\documentclass[10pt,journal,epsfig,twoside]{IEEEtran}
\usepackage{graphicx}
\usepackage{subfigure,color}
\usepackage{amssymb}
\usepackage{cite,comment}
\usepackage{amsmath}
\usepackage{bm,comment}
\usepackage{algorithm}
\usepackage{algpseudocode}
\makeatletter

\newcommand{\Rmnum}[1]{\expandafter\@slowromancap\romannumeral #1@}

\makeatother

\newtheorem{lemma}{\underline{Lemma}}

\begin{document}
\title{Cooperative NOMA for Downlink Asymmetric Interference Cancellation}
\author{Weidong Mei and Rui Zhang, \IEEEmembership{Fellow, IEEE}
\thanks{Manuscript received November 27, 2019; accepted February 11, 2020. The associate editor coordinating the review of this article and approving it for publication was W. Hamouda. {\it (Corresponding author: Weidong Mei.)}}
\thanks{W. Mei is with the NUS Graduate School for Integrative Sciences and Engineering, National University of Singapore, Singapore 119077, and also with the Department of Electrical and Computer Engineering, National University of Singapore, Singapore 117583 (e-mail: wmei@u.nus.edu).}
\thanks{R. Zhang is with the Department of Electrical and Computer Engineering, National University of Singapore, Singapore 117583 (e-mail: elezhang@nus.edu.sg).}}
\markboth{IEEE WIRELESS COMMUNICATIONS LETTERS}
{MEI and ZHANG: COOPERATIVE NOMA FOR DOWNLINK ASYMMETRIC INTERFERENCE CANCELLATION}
\maketitle

\begin{abstract}
This letter advances the non-orthogonal multiple access (NOMA) technique for cellular downlink co-channel interference mitigation, via exploiting the (limited) cooperation among base stations (BSs). Specifically, we consider a simplified but practically relevant scenario of two co-channel cells with {\it \textbf{asymmetric interference}}, i.e., only the user in one cell receives the strong interference from the BS in the other cell. To mitigate such interference, we propose a new {\it \textbf{cooperative NOMA}} scheme, where the interfered user's serving BS sends a superposed signal comprising both the desired message and the co-channel user's message (shared by the interfering BS). The co-channel user's signal is aimed to add constructively with the interfering BS's signal at the interfered user's receiver so that the combined interference with enhanced power can be effectively decoded and cancelled. This thus leads to a new problem on how to optimally allocate the transmit power for the two superposed messages. We provide the closed-form solution to this problem and investigate the conditions under which the performance of the proposed scheme is superior over the existing schemes.
\end{abstract}

\section{Introduction}
Thanks to its ability to realize massive connectivity, low latency and high spectral efficiency in wireless communications, non-orthogonal multiple access (NOMA) technique has been recognized as a key enabler for future cellular networks. As such, NOMA has drawn a great deal of attention from both academia and industry (see, e.g., \cite{islam2017power,ding2017survey,3GPP36859} and the references therein). However, most of the existing studies on NOMA are limited to the single-cell setup, while only a handful of works have recently addressed the more challenging multi-cell scenario (see e.g., \cite{choi2014non,tian2016performance,ali2018downlink,shin2016coordinated}). For multi-cell NOMA, inter-cell interference (ICI) is a major issue as it makes the successive interference cancellation (SIC) design far more complicated as compared to the single-cell case without ICI, especially for cell-edge user equipments (UEs) that generally suffer strong co-channel interference from other cells.

To resolve the above issue, NOMA has been combined with various interference mitigation techniques such as ICI coordination (ICIC) and cooperative multi-point (CoMP), generally referred to as cooperative NOMA, to reap its benefits as in the single-cell system. Specifically, cooperative NOMA involves multiple BSs to serve the cell-edge UEs at the same time by leveraging the message sharing among cooperating BSs. For example, in \cite{choi2014non}, the author proposed a coordinated superposition coding scheme in a two-cell downlink network, where a cell-edge UE is served by two BSs via Alamouti code. To reduce the complexity of multi-user NOMA, an opportunistic NOMA scheme was proposed in \cite{tian2016performance}, where each cell-edge UE is allowed to select its own preferred set of serving BSs. Furthermore, the work \cite{ali2018downlink} developed a general cooperation model with coexisting CoMP and non-CoMP UEs, and NOMA is applied at each BS to schedule their communications over the same resource block (RB). The authors in \cite{shin2016coordinated} proposed two interference alignment-based cooperative NOMA schemes so as to completely eliminate the ICI suffered by cell-edge UEs.

\begin{figure}[!t]
\centering
\subfigure[UE1 and UE2 are both terrestrial UEs.]{\includegraphics[width=0.45\textwidth]{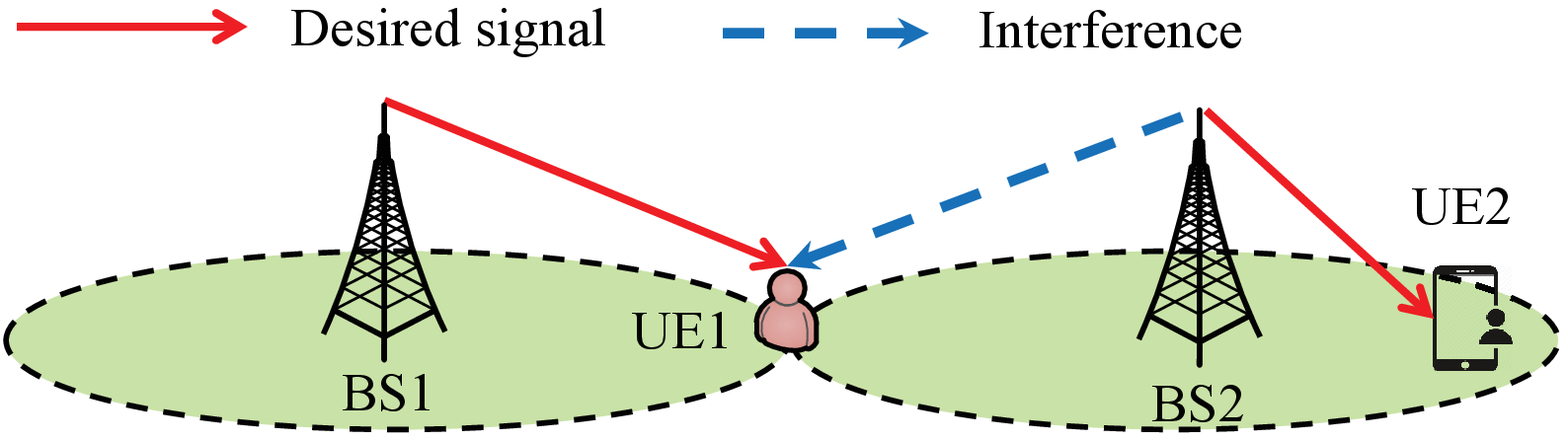}}
\subfigure[UE1 is a UAV UE and UE2 is a terrestrial UE.]{\includegraphics[width=0.45\textwidth]{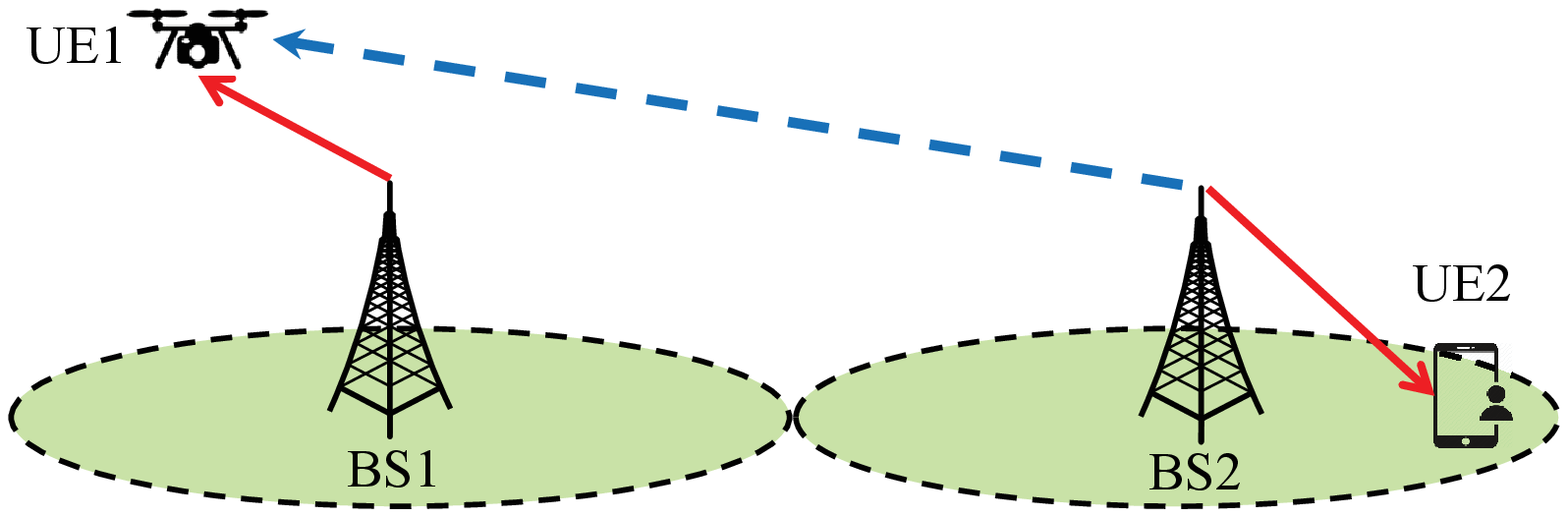}}
\caption{Downlink communication with asymmetric co-channel interference.}\label{down}\vspace{-12pt}
\end{figure}
As shown in Fig.\,\ref{down}, we consider in this letter a simplified two-cell system with {\it asymmetric interference}, where only the downlink transmission from BS1 to UE1 is strongly interfered by that from BS2 to UE2. This scenario can occur in practice, e.g., when UE1 is at the cell edges of both BS1 and BS2 and the distance between BS1 and UE2 is much longer than that between BS1/BS2 and UE1 (see Fig.\,\ref{down}(a)). Alternatively, if UE1 is an unmanned aerial vehicle (UAV)\cite{cooperative2019mei}, then it suffers much stronger ground-to-air interference from the co-channel BS2 as compared to the terrestrial interference caused by BS1 to UE2, even if UE1 (the UAV) is much closer to its serving BS (BS1) than the interfering BS (BS2), as shown in Fig.\,\ref{down}(b). This is because for high-altitude UAVs, their channels with ground BSs are dominated by line-of-sight (LoS) propagation\cite{3GPP36777}, and thus suffer less path-loss, shadowing and multi-path fading as compared to the typical terrestrial channels (e.g., that between BS1 and UE2 in Fig.\,\ref{down}) with rich scatterers\footnote{Please refer to \cite{cooperative2019mei} and \cite{mei2019uplink,liu2018multi,cellular2018mei} for more details on the aerial-ground interference mitigation techniques for the downlink and uplink cellular UAV communications, respectively.}. In both the above scenarios, despite that UE1 can perform SIC to subtract the interference from BS2, its achievable rate can be limited for satisfying the rate demand of UE2 due to the comparable signal and interference links from BS1 and BS2, respectively. To improve the rate performance of UE1 without affecting the co-channel transmission for UE2, we propose a new cooperation scheme by exploiting only one-sided message sharing from BS2 to BS1 (as opposed to the two-sided message sharing between the two BSs required by CoMP-based NOMA as in \cite{choi2014non,tian2016performance,ali2018downlink,shin2016coordinated}). Specifically, BS2 via the backhaul link shares the data symbol for UE2 with BS1, which then transmits a superposed signal comprising both UE1's and UE2's data symbols. As a result, UE2's signal can be added destructively or constructively with BS2's interference at UE1's receiver to suppress it (for decoding UE1's signal directly) or enhance it (for decoding and cancelling UE2's signal first before decoding UE1's signal), respectively. In both cases, BS1's transmit power allocations for UE1's and UE2's signals need to be optimized accordingly to maximize the achievable rate of UE1. For the former case, the optimal power allocation has been derived in our prior work \cite{cooperative2019mei}, while in this letter we address this problem in the latter case. Note that in this case, the proposed cooperative NOMA resembles the conventional NOMA \cite{islam2017power,ding2017survey,3GPP36859} in the sense that BS1 sends a superposed signal of both UEs, and the near UE (UE1) applies SIC to decode and cancel the interference due to the far UE (UE2). However, a key difference between them lies in that the transmitted UE2's signal in our scheme is not intended for UE2 (as in conventional NOMA), but instead for enhancing the combined interference at UE1's receiver to facilitate its SIC. We compare the proposed cooperative NOMA scheme with other existing schemes both analytically and numerically to characterize the conditions under which it gives superior rate performance.

{\it Notations:} For a complex number $s$, $\lvert s \rvert$ denotes its amplitude, $\angle s$ denotes its phase, and $s \sim \mathcal{CN}(\mu,\sigma^2)$ means that it is a circularly symmetric complex Gaussian (CSCG) random variable with mean $\mu$ and variance $\sigma^2$. ${\mathbb E}[\cdot]$ denotes the expected value of random variables.

\section{System Model and Problem Formulation}
As shown in Fig.\,\ref{down}, we consider a two-cell setting\footnote{This is for the convenience of illustrating the proposed scheme, while we will leave the extension to the general multi-cell system for our future work.}, where two BSs (BS1 and BS2) serve their respective UEs (UE1 and UE2) in the downlink over the same time-frequency RB. We assume that each BS employs an antenna array with fixed directional gain pattern, while each UE has a single antenna. We consider the asymmetric interference scenario as explained in Section \Rmnum{1}, where UE1 receives the strong co-channel interference from BS2, while the interference from BS1 to UE2 is negligible and thus ignored. It is assumed that the downlink transmission from BS2 to UE2 has already started over the considered RB before BS1 serves UE1 using the same RB. As such, we consider that BS2 cannot change its transmission to UE2 and thus the CoMP-based cooperative NOMA\cite{choi2014non,tian2016performance,ali2018downlink,shin2016coordinated} is not applicable. Nonetheless, BS2 can help BS1 mitigate its interference to UE1 by sharing some useful information via their backhaul link (e.g., the X2 interface in Long Term Evolution (LTE)\cite{dahlman20134g}), such as UE2's data symbol and the channel gain from it to UE1, which thus enables our proposed cooperative NOMA scheme. For ease of reference, the main symbols used in this letter are listed in Table \ref{variable}.\vspace{-9pt}
\begin{table}[t]
\centering
\caption{List of Main Symbols}\label{variable}
\begin{tabular}{|c|l|}
\hline
Symbol & Description\\
\hline
$P$ & Maximum transmit power of BS1 \\
\hline
$Q$ & Maximum transmit power of BS2 \\
\hline
$h_1$ & Baseband equivalent channel from BS1 to UE1 \\
\hline
$h_2$ & Baseband equivalent channel from BS2 to UE1 \\
\hline
$x_1$ & Complex data symbol for UE1 \\
\hline
$x_2$ & Complex data symbol for UE2 \\
\hline
$\sigma^2$ & UE1's receiver noise power \\
\hline
$\gamma^\star_{1,\text{S}i}$ & UE1's maximum SINR achievable by scheme $i, i=1,2,3,4$ \\
\hline
$w_1$ & Complex weight for transmitting $x_1$ by BS1\\
\hline
$w_2$ & Complex weight for transmitting $x_2$ by BS1\\
\hline
$v_1$ & Amplitude of of $w_1$, i.e., $\lvert w_1 \rvert$\\
\hline
$v_2$ & Amplitude of of $w_2$, i.e., $\lvert w_2 \rvert$\\
\hline
$\rho$ & Minimum SINR required for decoding UE2's message\\
\hline
\end{tabular}
\vspace{-12pt}
\end{table}

\subsection{Conventional Schemes}\label{conv.sc}
First, we characterize the achievable signal-to-interference-plus-noise ratio (SINR) of UE1 via the conventional single-cell schemes by BS1 without BS2's cooperation. Two schemes are considered, with or without SIC applied at UE1. Let $h_1$ be the complex-valued baseband equivalent channel gain from BS1 to UE1, and $h_2$ be that from BS2 to UE1. Let $P$ and $Q$ denote the maximum transmit power of BS1 and BS2, respectively. Then, the received signal at UE1 can be expressed as
\begin{equation}
y_1=\sqrt{P_1}h_1x_1+\sqrt{Q}h_2x_2+z_1,
\end{equation}
where $P_1 \le P$ denotes the transmit power of BS1, $x_1$ and $x_2$ denote the complex-valued data symbols for UE1 and UE2 with ${\mathbb E}[{\lvert{x_1}\rvert}^2] = 1$ and ${\mathbb E}[{\lvert{x_2}\rvert}^2] = 1$, respectively, and $z_1 \sim \mathcal{CN}(0,\sigma^2)$ denotes UE1's receiver noise with $\sigma^2$ denoting the power.

\textbf{\underline{Scheme 1}:} If SIC is not implemented at UE1, then the co-channel interference $\sqrt{Q}h_2x_2$ is treated as Gaussian noise at its receiver. In this case, BS1 should transmit at its full power, i.e., $P_1=P$, to maximize UE1's receive SINR, which can be expressed as
\begin{equation}\label{sinr1}
\gamma^\star_{1, \text{S1}} = \frac{P\lvert h_1 \rvert^2}{\sigma^2 + Q\lvert h_2 \rvert^2}.
\end{equation}

\textbf{\underline{Scheme 2}:} On the other hand, if UE1 first decodes UE2's message and then subtracts it, UE1 will be free of co-channel interference. As a result, its receive SINR is given by
\begin{equation}\label{sinr2}
\gamma_{1,\text{S2}} = \frac{P_1\lvert h_1 \rvert^2}{\sigma^2}.
\end{equation}
Note that to successfully cancel UE2's signal, its receive SINR at UE1 is given by
\begin{equation}\label{sinr2}
\gamma_{2,\text{S2}} = \frac{Q\lvert h_2 \rvert^2}{\sigma^2 + P_1\lvert h_1 \rvert^2}.
\end{equation}

As a result, the maximum receive SINR of UE1 under scheme 2 can be obtained by solving the following optimization problem
\begin{equation}\label{op1}
\text{(P-S2)}\;\mathop {\max}\limits_{0 \le P_1 \le P}\; \gamma_{1,\text{S2}},\quad\text{s.t.}\; \gamma_{2,\text{S2}} \ge \rho,
\end{equation}
where $\rho$ denotes the minimum SINR required for decoding UE2's message (say, at UE2' receiver), which is assumed to be given and fixed. Obviously, (P-S2) is feasible if and only if $\frac{Q\lvert h_2 \rvert^2}{\sigma^2} \ge \rho$. Assuming that (P-S2) is feasible, its optimal solution, denoted as $P_1^\star$, can be easily shown to be
\begin{equation}
P_1^\star=
\begin{cases}
\frac{Q{\lvert h_2 \rvert}^2}{\rho{\lvert h_1 \rvert}^2} - \frac{\sigma ^2}{{\lvert h_1 \rvert}^2}, &\text{if}\;\frac{Q\lvert h_2 \rvert^2}{\sigma^2 + P\lvert h_1 \rvert^2} \le \rho\\
P, &\text{otherwise.}
\end{cases}
\end{equation}
Accordingly, its optimal value, denoted as $\gamma^\star_{1,\text{S2}}$, is given by
\begin{equation}\label{sinr3}
\gamma^\star_{1,\text{S2}}=\frac{P_1^\star\lvert h_1 \rvert^2}{\sigma^2}=
\begin{cases}
\frac{Q{\lvert h_2 \rvert}^2}{\rho\sigma ^2} - 1, &\text{if}\;\frac{Q\lvert h_2 \rvert^2}{\sigma^2 + P\lvert h_1 \rvert^2} \le \rho\\
\frac{P\lvert h_1 \rvert^2}{\sigma^2}, &\text{otherwise.}
\end{cases}
\end{equation}
\vspace{-12pt}

\subsection{Cooperation Schemes}
Next, we consider the case where BS2 cooperatively sends $h_2$ and $x_2$ to BS1 to facilitate the interference cancellation at UE1. With $x_2$ available at BS1, it transmits the superposition of $x_1$ and $x_2$, i.e., $w_1x_1 + w_2x_2$, where $w_1$ and $w_2$ denote the complex weights. To satisfy the power constraint at BS1, it must hold that $\lvert w_1 \rvert^2  + \lvert w_2 \rvert^2 \le P$. Then, the received signal at UE1 becomes
\begin{equation}\label{superpose}
y_1=h_1w_1x_1 + (h_1w_2+\sqrt{Q}h_2)x_2+z_1.
\end{equation}

Based on (\ref{superpose}), we introduce two interference cancellation schemes for UE1, depending on whether $h_1w_2$ is designed to be in- or out-of-phase with the interference channel gain $h_2$.

\textbf{\underline{Scheme 3}\cite{cooperative2019mei}:} If $h_1w_2$ is designed to be opposite to $h_2$, i.e., $\angle w_2= \angle h_2-\angle h_1 + \pi$, the interference due to $x_2$ at UE1 can be suppressed, and its receive SINR can be improved as compared to (\ref{sinr1}), which is given by
\begin{equation}\label{sinr5}
\gamma_{1,\text{S3}}=\frac{\lvert h_1 \rvert^2\lvert w_1 \rvert^2}{\sigma ^2 + (\lvert h_2 \rvert\sqrt{Q} -\lvert h_1 \rvert \lvert w_2 \rvert)^2}.
\end{equation}
For convenience, let $v_1 \triangleq \lvert w_1 \rvert$ and $v_2 \triangleq \lvert w_2 \rvert$ be the amplitude of the complex weights $w_1$ and $w_2$, respectively. Then the problem for maximizing (\ref{sinr5}) can be formulated as
\begin{align}
\text{(P-S3)}\mathop {\max}\limits_{v_1,v_2 \ge 0}&\; \frac{\lvert h_1 \rvert^2v^2_1}{\sigma ^2 + (\lvert h_2 \rvert\sqrt{Q} -\lvert h_1 \rvert v_2)^2}\nonumber\\
\text{s.t.}\;\;&v_1^2  + v_2^2 \le P.\label{op2}
\end{align}
Notice that with $v_1 = 0$ in (P-S3), scheme 3 reduces to scheme 1. Consequently, the solution to (P-S3) should generally yield a higher receive SINR for UE1 than scheme 1 without BS2's cooperation.

From \cite{cooperative2019mei}, the optimal solution to (P-S3), denoted by $(v_1^\star,v_2^\star)$, is given by
\begin{equation}
\begin{split}
v_2^\star&=\frac{X - \sqrt {X^2 - 4{\lvert h_1 \rvert}^2{\lvert h_2 \rvert}^2PQ}}{2{\lvert h_1 \rvert}{\lvert h_2 \rvert}\sqrt Q},\\
\;v_1^\star&=\sqrt {P - v_2^{\star 2}},
\end{split}
\end{equation}
where $X \triangleq \sigma^2 + {\lvert h_1 \rvert}^2P + {\lvert h_2 \rvert}^2Q$. Moreover, UE1's maximum receive SINR, denoted as $\gamma^\star_{1,\text{S3}}$, is given by
\begin{equation}\label{itc}
\gamma^\star_{1,\text{S3}}=\frac{-Y + \sqrt {Y^2 + 4\sigma^2P{{\lvert h_1 \rvert}^2}}}{2\sigma^2},
\end{equation}
where $Y \triangleq \sigma ^2 + Q{\lvert h_2 \rvert}^2 - P{\lvert h_1 \rvert}^2$. It is worth noting that SIC is not applied at UE1's receiver in this scheme. Moreover, it is shown in \cite{cooperative2019mei} that $\gamma^\star_{1,\text{S3}}$ monotonically increases with $P{\lvert h_1 \rvert}^2 - Q{\lvert h_2 \rvert}^2$. This implies that if the interference power $Q{\lvert h_2 \rvert}^2$ becomes stronger (relative to the desired signal power $P{\lvert h_1 \rvert}^2$), scheme 3 achieves lower SINR and thus becomes less effective.

\textbf{\underline{Scheme 4}:} Note that in scheme 2, the use of SIC for canceling UE2's signal at UE1 limits the achievable rate of UE1, especially when the desired signal power $P{\lvert h_1 \rvert}^2$ becomes comparable with the interference power $Q{\lvert h_2 \rvert}^2$ or $\rho$ is large (i.e., when the first case in (\ref{sinr3}) is likely to be true). To improve over scheme 2, a new cooperative NOMA scheme, referred to as scheme 4, is proposed in this letter, where $h_1w_2$ is designed to be in-phase with the interference channel gain $h_2$, i.e., $\angle w_2 = \angle h_2 - \angle h_1$, for enhancing the combined interference due to UE2's signal so as to cancel it more effectively by SIC at UE1's receiver.

As a result, with scheme 4, UE1's and UE2's achievable SINRs with SIC can be expressed as
\begin{equation}
\begin{split}
\gamma_{1,\text{S4}}&=\frac{\lvert h_1 \rvert^2v_1^2}{\sigma ^2},\\
\gamma_{2,\text{S4}}&=\frac{(\lvert h_1 \rvert v_2 + \sqrt{Q}\lvert h_2 \rvert)^2}{\sigma ^2 + \lvert h_1 \rvert^2 v_1^2},
\end{split}
\end{equation}
respectively. To ensure that UE2's signal can be decoded, the following inequality should be met, i.e., $\gamma_{2,\text{S4}} \ge \rho$.

The new power allocation problem for maximizing $\gamma_{1,\text{S4}}$ is thus formulated as
\begin{subequations}\label{op3}
\begin{align}
\text{(P-S4)} \mathop {\max}\limits_{v_1,v_2 \ge 0}\; &\frac{\lvert h_1 \rvert^2v_1^2}{\sigma ^2} \nonumber\\
\text{s.t.}\;\;&\frac{(\lvert h_1 \rvert v_2 + \sqrt{Q}\lvert h_2 \rvert)^2}{\sigma ^2 + \lvert h_1 \rvert^2 v_1^2} \ge \rho, \label{op3a}\\
&v_1^2 + v_2^2 \le P.\label{op3b}
\end{align}
\end{subequations}
Notice that with $v_2 = 0$ in problem (P-S4), (P-S4) reduces to (P-S2). Consequently, the proposed cooperative NOMA scheme generally yields a higher SINR or achievable rate for UE1 than the conventional NOMA (scheme 2) without BS2's cooperation.\vspace{-6pt}

\section{Optimal Solution and Performance Comparison}
In this section, we first derive the optimal solution to (P-S4) which achieves the maximum receive SINR of UE1 by our proposed scheme (scheme 4). Then, we compare the performance of the proposed scheme with that of scheme 3 to reveal the conditions under which the proposed scheme achieves superior performance.
\vspace{-8pt}

\subsection{Optimal Solution to (P-S4)}\label{opt.sol}
It is easy to show, by contradiction, that the constraint (\ref{op3b}) must hold with equality at the optimality of (P-S4), i.e., $v_1^2+v_2^2=P$. Otherwise, we can construct a new solution $(\hat v_1, \hat v_2)=(\sqrt{v_1^2+\delta},\sqrt{v_2^2+\rho\delta})$ with $\delta=(P-v_1^2-v_2^2)/(\rho+1)$. Obviously, we have $\hat v_1^2+\hat v_2^2=P$ and
\begin{equation}\label{ineq1}
\frac{(\lvert h_1 \rvert \hat v_2 + \sqrt{Q}\lvert h_2 \rvert)^2}{\sigma ^2 + \lvert h_1 \rvert^2 \hat v_1^2}=\frac{(\lvert h_1 \rvert v_2 + \sqrt{Q}\lvert h_2 \rvert)^2+\lvert h_1 \rvert^2\rho\delta+2Z}{\sigma^2+\lvert h_1 \rvert^2 v_1^2+\lvert h_1 \rvert^2 \delta},
\end{equation}
with $Z\triangleq\sqrt{Q}\lvert h_1 \rvert\lvert h_1 \rvert(\hat v_2-v_2)>0$. Since $(v_1, v_2)$ is a feasible solution to (P-S4), it should satisfy the constraint (\ref{op3a}), i.e.,
$(\lvert h_1 \rvert v_2 + \sqrt{Q}\lvert h_2 \rvert)^2 \ge \rho(\sigma^2+\lvert h_1 \rvert^2 v_1^2)$. Moreover, as $\lvert h_1 \rvert^2\rho\delta+2Z>\lvert h_1 \rvert^2\rho\delta$, it follows that the right-hand side (RHS) of (\ref{ineq1}) is greater than $\rho$. Hence, $(\hat v_1, \hat v_2)$ is a feasible solution to (P-S4). However, since $\hat v_1 > v_2$, this new solution yields a larger objective value of (P-S4) than $(v_1,v_2)$. This contradicts the presumption, and thus $v_1^2+v_2^2=P$ must hold at the optimality of (P-S4). By substituting $v_1^2=P-v_2^2$ into (P-S4), we obtain the following equivalent problem with only a single variable $v_2$, i.e.,
\begin{equation}\label{op4}
\mathop {\max}\limits_{0 \le v_2 \le \sqrt{P}}\; P-v_2^2, \quad\text{s.t.}\;\;\frac{(\lvert h_1 \rvert v_2 + \sqrt{Q}\lvert h_2 \rvert)^2}{\sigma ^2 + \lvert h_1 \rvert^2(P-v_2^2)} \ge \rho,
\end{equation}
where the constant term $\lvert h_1 \rvert^2/\sigma^2$ is omitted in the objective function. Since $v_2 \ge 0$, the above problem is equivalent to
\begin{equation}\label{op5}
\mathop {\min}\limits_{0 \le v_2 \le \sqrt{P}}\; v_2, \quad \text{s.t.}\;\; F(v_2) \ge \rho,
\end{equation}
where $F(v_2) \triangleq \frac{(\lvert h_1 \rvert v_2 + \sqrt{Q}\lvert h_2 \rvert)^2}{\sigma ^2 + \lvert h_1 \rvert^2(P-v_2^2)}$.

It is easy to verify that as $v_2$ increases, the numerator and the denominator of $F(v_2)$ increase and decrease, respectively. As such, $F(v_2)$ is a monotonically increasing function of $v_2$. It then follows that problem (\ref{op5}) is feasible if and only if $F(\sqrt{P}) \ge \rho$, which can be shown equivalent to
\begin{equation}\label{cond1}
\frac{(\sqrt{P}\lvert h_1 \rvert + \sqrt{Q}\lvert h_2 \rvert)^2}{\sigma^2} \ge \rho.
\end{equation}
Moreover, if $F(0)=\frac{Q\lvert h_2 \rvert^2}{\sigma^2 + P\lvert h_1 \rvert^2} \ge \rho$, i.e., the optimal solution to problem (\ref{op5}) is $v_2=0$, scheme 4 becomes equivalent to scheme 2. Finally, if $F(0) < \rho \le F(\sqrt{P})$, the optimal solution to (P2) should be the solution to the equation $F(v_2) = \rho$, or equivalently, the quadratic equation $G(v_2)=0$, where
\begin{equation}
G(v_2) \!=\! (1+\rho)\lvert h_1 \rvert^2v_2^2+2\sqrt{Q}\lvert h_1 \rvert\lvert h_2 \rvert v_2+Q\lvert h_2 \rvert^2\!-\!\rho\sigma^2\!-\!P\rho\lvert h_1 \rvert^2\!.
\end{equation}
Since $Q\lvert h_2 \rvert^2-\rho\sigma^2-P\rho\lvert h_1 \rvert^2 = (F(0)-\rho)(\sigma^2+P\lvert h_1 \rvert^2)<0$, the quadratic equation $G(v_2)=0$ only has a single positive root, which is the optimal solution to problem (\ref{op4}) and given by
\begin{equation}\label{optSol}
v_2^*=\frac{\sqrt A - \sqrt Q \lvert h_2 \rvert}{\lvert h_1 \rvert(1 + \rho)},
\end{equation}
where $A \triangleq P\rho(1+\rho)\lvert h_1 \rvert^2+\rho(\rho+1)\sigma^2-Q\rho\lvert h_2 \rvert^2$.

Correspondingly, if scheme 4 is feasible, i.e., $\rho \le F(\sqrt{P})$, UE1's maximum receive SINR can be expressed as
\begin{equation}\label{sinr6}
\gamma^\star_{1,{\text{S4}}}=
\begin{cases}
\frac{\lvert h_1 \rvert^2}{\sigma ^2}(P-v_2^{*2}), &\text{if}\;\rho > F(0)\\
\frac{P\lvert h_1 \rvert^2}{\sigma^2}, &\text{otherwise}.
\end{cases}
\end{equation}

It follows from (\ref{optSol}) that when the interference power $Q\lvert h_2 \rvert^2$ increases, the numerator of $v_2^*$ decreases. As such, $v_2^*$ monotonically decreases with $Q\lvert h_2 \rvert^2$. This implies that UE1's maximum receive SINR, as given in (\ref{sinr6}), is non-decreasing with $Q\lvert h_2 \rvert^2$. This is in a sharp contrast to scheme 3 for which UE1's maximum receive SINR, i.e., $\gamma^\star_{1,{\text{S3}}}$ in (\ref{itc}),  decreases with $Q\lvert h_2 \rvert^2$. The above observations imply that our proposed scheme (scheme 4) is more advantageous over scheme 2 or 3 when $\rho$ is larger or the interference power $Q\lvert h_2 \rvert^2$ is larger, respectively. \vspace{-6pt}

\subsection{Performance Comparison}\label{perf.comp}
Since scheme 2 is a special case of scheme 4, while scheme 1 is a special case of scheme 3, it suffices to compare the performance of our proposed scheme 4 with that of scheme 3 analytically, as pursued in this subsection. For convenience, we define $\alpha \triangleq \frac{P\lvert h_1 \rvert^2}{\sigma^2}$ and $\beta \triangleq \frac{Q\lvert h_2 \rvert^2}{\sigma^2}$.

To this end, we compare (\ref{itc}) with (\ref{sinr6}). Firstly, if $\rho \le F(0)$, scheme 4 (or scheme 2) outperforms scheme 3 as $\gamma^\star_{1,\text{S4}}=\frac{P\lvert h_1 \rvert^2}{\sigma^2} \ge \gamma^\star_{1,\text{S3}}$. Secondly, if $F(0) \le \rho \le F(\sqrt{P})$, it can be shown that $\gamma^\star_{1,\text{S4}} \ge \gamma^\star_{1,\text{S3}}$ if $v_2^* \le \sqrt {P\xi}$, where $\xi = \frac{(1 +\alpha + \beta)-\sqrt{{(1 - \alpha + \beta )}^2 + 4\alpha}}{2\alpha}$. Since $v_2^*$ is the unique positive root of the quadratic equation $G(v_2) = 0$, the above inequality holds if and only if $G\left(\sqrt {P\xi}\right) \ge 0$, which, after some manipulations, can be shown equivalent to
\begin{equation}\label{cond6}
\rho \le \frac{(1+\alpha+3\beta)-\sqrt{(1-\alpha+\beta)^2+4\alpha}+2\sqrt W}{1+\alpha-\beta+\sqrt{(1-\alpha+\beta)^2+4\alpha}},
\end{equation}
where $W=2\beta(1+\alpha+\beta-\sqrt{(1-\alpha+\beta)^2 + 4\alpha})$.

By combining the results in the above two cases, it follows that scheme 4 yields a better performance than scheme 3 if the condition (\ref{cond6}) is met. Since $\gamma^\star_{1,\text{S4}}$ and $\gamma^\star_{1,\text{S3}}$ monotonically increase and decrease with the interference power $Q\lvert h_2 \rvert^2$, respectively, the threshold given in the RHS of (\ref{cond6}) must monotonically increase with $\beta$ or the interference power at UE1, $Q\lvert h_2 \rvert^2$, which is in accordance with our previous discussion at the end of Section \ref{opt.sol}, as will be also shown via numerical results in the next section.

\section{Numerical Results}
In this section, numerical results are provided to evaluate the performance of the proposed cooperative NOMA scheme (scheme 4), as compared to the benchmark schemes 1, 2 and 3. We consider a cellular-connected UAV for UE1, while UE2 is a terrestrial user. Unless otherwise specified, the simulation settings are as follows. The bandwidth is set to 180 kHz, which is equal to the width of a time-frequency RB in LTE\cite{dahlman20134g}. The carrier frequency $f_c$ is $2$ GHz, and the noise power spectrum density at UE1's receiver is $-164$ dBm/Hz. The height of BSs is set to be 25 in meter (m). The altitude of the UAV is fixed as 200 m. The horizontal distance between the UAV and BS1 (BS2) is 0.92 km (2.88 km). The BS antenna elements are placed vertically with half-wavelength spacing and electrically steered with 10-degree downtilt angle. The UAV-BS channels follow the probabilistic LoS channel model based on the urban macro scenario in \cite{3GPP36777}. The transmit power of BS1 is set to be $P=20$ dBm.

\begin{figure}[htb]
\centering
\includegraphics[width=3.2in]{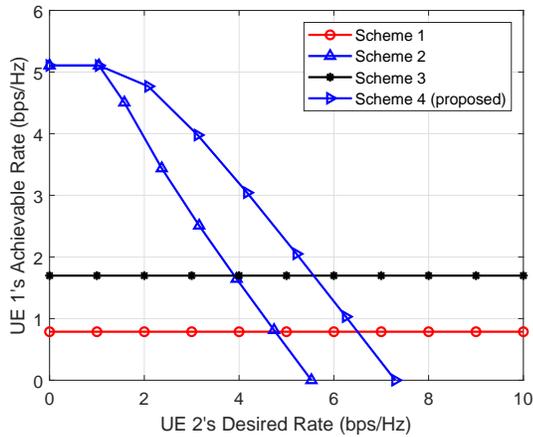}
\DeclareGraphicsExtensions.
\caption{UE1's achievable rate versus UE2's given rate.}\label{IERate_QoS}
\vspace{-8pt}
\end{figure}
Fig.\,\ref{IERate_QoS} shows UE1's achievable rate (defined as $\log_2(1+\text{SINR})$ in bits per second per Hertz (bps/Hz), where SINR denotes the maximum achievable SINR in each scheme) by different schemes versus UE2's given rate, $\log_2(1+\rho)$. The transmit power of BS2 is assumed to be identical to that of BS1, i.e., $P=Q=20$ dBm. It is observed that the performance of schemes 2 and 4 decreases with increasing UE2's rate or $\rho$, since more transmit power needs to be allocated for transmitting UE2's message by BS1 in order to cancel its (combined) interference at UE1 by SIC. In contrast, without the need of applying SIC at UE1's receiver to cancel UE2's interference, the performance of schemes 1 and 3 is observed to be unaffected by UE2's rate. In addition, it is observed that the proposed scheme 4 significantly outperforms schemes 1 and 3 when UE2's rate is not high. Moreover, the performance gap between schemes 2 and 4 is observed to be enlarged as UE2's rate increases, which is consistent with our discussion at the end of Section \ref{opt.sol}.

\begin{figure}[htb]
\centering
\includegraphics[width=3.2in]{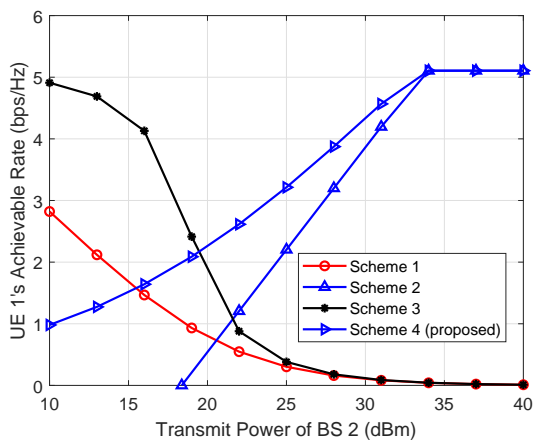}
\DeclareGraphicsExtensions.
\caption{UE1's achievable rate versus BS2's transmit power.}\label{IERate_Intf}
\vspace{-8pt}
\end{figure}
Next, we plot UE1's achievable rate versus BS2's transmit power $Q$ in Fig.\,\ref{IERate_Intf} by fixing $\log_2(1+\rho)=5$ bps/Hz. It is observed that UE1's achievable rates by schemes 1 and 3 quickly diminish as $Q$ increases, due to the increasing (residual) co-channel interference. In contrast, UE1's achievable rates by schemes 2 and 4 increase with $Q$ or the interference power and finally converge to the same maximum value when UE2's rate can be satisfied even without BS2's cooperation, i.e., $\log_2(1+\frac{P\lvert h_1 \rvert^2}{\sigma^2})$, corresponding to the second case of (\ref{sinr3}) and (\ref{sinr6}). It is also observed that scheme 4 outperforms scheme 3 when $Q$ or the interference power at UE1 is sufficiently large, as analytically shown in Section \ref{perf.comp}.

\section{Conclusions}
This letter proposes a new cooperative NOMA scheme for cellular downlink to resolve the strong asymmetric interference issue. The key difference from the conventional NOMA lies in the new superposition signal design for the purpose of enhancing the co-channel interference at the receiver to facilitate SIC. It is shown both analytically and numerically that the proposed scheme significantly outperforms the conventional NOMA with SIC (scheme 2) when the co-channel interference is comparable to the desired signal in power, as well as the existing interference transmission and cancellation (ITC) scheme without SIC (scheme 3)\cite{cooperative2019mei} when the co-channel interference is strong. Both scenarios may practically occur in cellular networks (e.g., for cellular-connected UAVs).

\bibliography{Coop_NOMA}

\begin{thebibliography}{10}
\providecommand{\url}[1]{#1}
\csname url@samestyle\endcsname
\providecommand{\newblock}{\relax}
\providecommand{\bibinfo}[2]{#2}
\providecommand{\BIBentrySTDinterwordspacing}{\spaceskip=0pt\relax}
\providecommand{\BIBentryALTinterwordstretchfactor}{4}
\providecommand{\BIBentryALTinterwordspacing}{\spaceskip=\fontdimen2\font plus
\BIBentryALTinterwordstretchfactor\fontdimen3\font minus
  \fontdimen4\font\relax}
\providecommand{\BIBforeignlanguage}[2]{{%
\expandafter\ifx\csname l@#1\endcsname\relax
\typeout{** WARNING: IEEEtran.bst: No hyphenation pattern has been}%
\typeout{** loaded for the language `#1'. Using the pattern for}%
\typeout{** the default language instead.}%
\else
\language=\csname l@#1\endcsname
\fi
#2}}
\providecommand{\BIBdecl}{\relax}
\BIBdecl

\bibitem{islam2017power}
S.~R. Islam, N.~Avazov, O.~A. Dobre, and K.-S. Kwak, ``Power-domain
  non-orthogonal multiple access ({NOMA}) in 5{G} systems: Potentials and
  challenges,'' \emph{{IEEE} Commun. Surveys Tuts.}, vol.~19, no.~2, pp.
  721--742, 2nd Quart. 2017.

\bibitem{ding2017survey}
Z.~Ding, X.~Lei, G.~K. Karagiannidis, R.~Schober, J.~Yuan, and V.~K. Bhargava,
  ``A survey on non-orthogonal multiple access for 5{G} networks: Research
  challenges and future trends,'' \emph{{IEEE} J. Sel. Areas Commun.}, vol.~35,
  no.~10, pp. 2181--2195, Oct. 2017.

\bibitem{3GPP36859}
\BIBentryALTinterwordspacing
{3GPP-TR-36.859}, ``Study on downlink multiuser superposition transmission
  ({MUST}) for {LTE},'' 2015, 3GPP technical report. [Online]. Available:
  \url{www.3gpp.org/dynareport/36859.htm}
\BIBentrySTDinterwordspacing

\bibitem{choi2014non}
J.~Choi, ``Non-orthogonal multiple access in downlink coordinated two-point
  systems,'' \emph{{IEEE} Commun. Lett.}, vol.~18, no.~2, pp. 313--316, Feb.
  2014.

\bibitem{tian2016performance}
Y.~Tian, A.~R. Nix, and M.~Beach, ``On the performance of opportunistic {NOMA}
  in downlink {CoMP} networks,'' \emph{{IEEE} Commun. Lett.}, vol.~20, no.~5,
  pp. 998--1001, May 2016.

\bibitem{ali2018downlink}
M.~S. Ali, E.~Hossain, A.~Al-Dweik, and D.~I. Kim, ``Downlink power allocation
  for {CoMP-NOMA} in multi-cell networks,'' \emph{{IEEE} Trans. Commun.},
  vol.~66, no.~9, pp. 3982--3998, Sep. 2018.

\bibitem{shin2016coordinated}
W.~Shin, M.~Vaezi, B.~Lee, D.~J. Love, J.~Lee, and H.~V. Poor, ``Coordinated
  beamforming for multi-cell {MIMO-NOMA},'' \emph{{IEEE} Commun. Lett.},
  vol.~21, no.~1, pp. 84--87, Jan. 2016.

\bibitem{cooperative2019mei}
\BIBentryALTinterwordspacing
W.~Mei and R.~Zhang, ``Cooperative downlink interference transmission and
  cancellation for cellular-connected {UAV}: A divide-and-conquer approach,''
  \emph{{IEEE} Trans. Commun.}, to be published. [Online]. Available:
  \url{https://arxiv.org/pdf/1906.00220.pdf}
\BIBentrySTDinterwordspacing

\bibitem{3GPP36777}
\BIBentryALTinterwordspacing
{3GPP-TR-36.777}, ``Study on enhanced {LTE} support for aerial vehicles,''
  2017, 3GPP technical report. [Online]. Available:
  \url{www.3gpp.org/dynareport/36777.htm}
\BIBentrySTDinterwordspacing

\bibitem{mei2019uplink}
W.~Mei and R.~Zhang, ``Uplink cooperative {NOMA} for cellular-connected
  {UAV},'' \emph{IEEE J. Sel. Topics Signal Process.}, vol.~13, no.~3, pp.
  644--656, Jun. 2019.

\bibitem{liu2018multi}
L.~Liu, S.~Zhang, and R.~Zhang, ``Multi-beam {UAV} communication in cellular
  uplink: Cooperative interference cancellation and sum-rate maximization,''
  \emph{{IEEE} Trans. Wireless Commun.}, vol.~18, no.~10, pp. 4679--4691, Oct.
  2019.

\bibitem{cellular2018mei}
W.~Mei, Q.~Wu, and R.~Zhang, ``Cellular-connected {UAV:} uplink association,
  power control and interference coordination,'' \emph{{IEEE} Trans. Wireless
  Commun.}, vol.~18, no.~11, pp. 5380--5393, Nov. 2019.

\bibitem{dahlman20134g}
E.~Dahlman, S.~Parkvall, and J.~Skold, \emph{{4G: LTE/LTE-advanced} for mobile
  broadband}.\hskip 1em plus 0.5em minus 0.4em\relax Oxford, UK: Academic
  press, 2013.

\end{thebibliography}
\bibliographystyle{IEEEtran}

\end{document}